\documentclass
{article}
\usepackage{algorithm}
\usepackage{algorithmic}
\usepackage{pdfsync}
\usepackage{graphicx}
\usepackage{subfigure}
\usepackage{pstricks}
\usepackage{animate}
\usepackage{pstricks-add}
\usepackage{multido}
\usepackage{color}
\usepackage{fancybox}
\usepackage{fancyvrb}
\usepackage{setspace}
\usepackage{amsmath}
\usepackage[colorlinks,citecolor=blue]{hyperref}

\begin{document}
\begin{center} 
\huge\textbf{Dynamical noncommutative quantum mechanics}
\end{center}

\begin{center}

\large\textbf{S. A. Alavi, S. Abbaspour}\\

\textit{Department of Physics, Hakim Sabzevari  University, P. O. Box 397, Sabzevar, Iran.}
\textit{}
 \end{center}
\begin{center}
\textbf{s.alavi@hsu.ac.ir ; alialavi@fastmail.us}
 \end{center}

\begin{abstract}
\emph{We study some basic  and interesting quantum mechanical systems  in dynamical noncommutative spaces in which the space- space  commutation relations are position dependent. It is observed  that the fundamental objects in the dynamical noncommutative space introduced here are stringlike.  We show  that the Stark effect can be employed to determine whether the noncommutativity of space is dynamical or non-dynamical. It appears  that unlike non-dynamical  case there is  a fundamental energy $\dfrac{\tau\hbar^{2}}{m}$  in this dynamical space.}
 \end{abstract}

\section{Introduction}\label{s1}
The idea of extension of noncommutativity to the coordinates was first suggested by Heisenberg as a possible solution for removing the infinite quantities of field theories. The renewed interest by physics community started to grow after the paper by Seiberg and Witten [1], see also [2]. Noncommutative quantum mechanics has received a wide attention in recent years and many physical problems have been studied in the framework of the noncommutative quantum mechanics.\\
The noncommutativity of the coordinates can be described by the following commutation relation:
 \begin{align}\label{p1}
[x_{\mu},x_{\nu}]=i\theta_{\mu \nu}
\end{align}
where  $\theta_{\mu \nu}$ is an anti-symmetric tensor. The simplest case corresponds to $\theta_{\mu \nu}$ being constant, which we call it non-dynamical or $\theta$-noncommutative spaces. In [3], the authors assumed $\theta_{\mu\nu}$ to be a function of position coordinates.\\
Recently [4] a generalization to dynamical( position dependent ) noncommutative spaces (DNCS) has been proposed in which the noncommutativity tensor is not constant but is position dependent and the following commutation relations  are  introduced  for a two dimensional dynamical noncommutative space :
 \begin{align}\label{p2}
[X,Y]&=i\theta (1+\tau Y^2)\ \ \ ; \ \ [X,P_x ]=i\hbar (1+\tau Y^2)\ \ \ ; \ \ [X,P_y ]=2i\tau Y(\theta P_y+\hbar X) \nonumber\\
[Y,P_y ]&=i \hbar (1+\tau Y^2 ) \  \ \ ; \ \ \  [Y,P_x ]=0    \ \ \ \ \ \ \ \ \ \ \ \ \ \ \ \ \ ; \ \ \  [P_x,P_y ]=0  
\end{align}
It is worth mentioning that by taking $\tau\rightarrow 0$, we recover the non-dynamical ($\theta$-noncommutative) commutation relations:
\begin{align}\label{g2}
 [x_0,y_0]&=i\theta \ \ \ \ \ \ ; \ \ \ \ \ [x_0,p_{x_0}]=i\hbar\ \ \ \ \ \ ; \ \ \ \ \ [x_0,p_{y_0}]=0
\nonumber\\
[y_0,p_{y_0}]&=i\hbar \ \ \ \ \ \ ; \ \ \ \ \ [y_0,p_{x_0}]=0 \ \ \ \ \ \ ; \ \ \ \ \ [p_{x_0},p_{y_0}]=0
\end{align}

The $X$ coordinate and the momentum in $Y$ direction, $P_y$ are not Hermitian so the Hamiltonian involving these variables will in general also not be Hermitian. But we may look for a similarity transformation i.e. a Dyson map $ \eta \ O\  \eta^{-1}=o=o^{\dag}$ which convert the non-Hermitian system into a Hermitian one. It is shown that the relevant Dyson map is $\eta=(1+\tau Y^2)^{-1/2}$, so the new Hermitian variables $x$, $y$, $p_x$ and $p_y$ can be stated in terms of $\theta$-noncommutative space variables as follows [4]:
\begin{align}\label{p6}
x=\eta X \eta^{-1}&=(1+\tau y_0^2)^{-1/2}(1+\tau y_0^2) x_0 (1+\tau
y_0^2)^{1/2}\nonumber\\&=(1+\tau y_0^2)^{1/2}x_0(1+\tau y_0^2)^{1/2}\nonumber\\
y=\eta Y \eta^{-1}&=(1+\tau y_0^2)^{-1/2}y_0 (1+\tau
y_0^2)^{1/2}=y_0 \nonumber \\
p_x=\eta P_x \eta^{-1}&=(1+\tau y_0^2)^{-1/2}p_{x_0} (1+\tau
y_0^2)^{1/2}=p_{x_0}\nonumber \\
p_y=\eta P_y \eta^{-1}&=(1+\tau y_0^2)^{-1/2}(1+\tau y_0^2) p_{y_0}
(1+\tau y_0^2)^{1/2}\nonumber\\&=(1+\tau y_0^2)^{1/2}p_{x_0}(1+\tau
y_0^2)^{1/2}
\end{align}
where zero index indicates the coordinates in $\theta$-noncommutative spaces.\\
These Hermitian dynamical noncommutative variables satisfy the following relations:
\begin{align}\label{p7}
[x,y]&=i\theta (1+\tau y^2)\ \ \ \ ; \ \ \ \   [x,p_x ]=i\hbar (1+\tau y^2) \ \ \ \ ; \ \ \ \   [p_x,p_y ]=0 
\nonumber\\
[x,p_y ]&=2i\tau y(\theta p_y+\hbar x)     \ \ \ \ ; \ \ \ \ [y,p_x ]=0 \ \ \ \ ; \ \ \ \       [y,p_y ]=i \hbar (1+\tau y^2 )
\end{align}
Using Bopp-shift one can relate the $\theta$-noncommutative variables to the variables of standard (conventional) commutative space:
\begin{equation}\label{p8}
x_{i_0}=x_{i_s}-\dfrac{\theta_{ij}}{2\hbar}p_{j_s}\ \ \ \ ,\ \ \  p_{i_0}=p_{i_s}\ \ \ \ ,\ \ \ \ i,j=x,y
\end{equation}
where $\theta_{ij}=\epsilon_{ijk}\theta_k$ and we take $\theta_3=\theta$ and the rest of the $\theta-$components to zero  [7].\\
An interesting point is that the minimal uncertainty of the coordinate $X$:
 \begin{equation}\label{p3}
\Delta X_{min}=\theta \sqrt{\tau}\sqrt{1+\tau \langle Y\rangle^2_{\rho}}
\end{equation}
leads to a minimal length for $X$ in a simultaneous $X$, $Y$ measurment, but there is no nonvanishing length in the $Y$-direction [4]. Here $\rho=\eta^{2}=(1+\tau Y^{2})^{-1}$, is the metric operator, and  $\langle Y\rangle_{\rho}$ is the expectation value of 
 operator $Y$  with respect to this metric.\\
This means that the fundamental objects  in this type of dynamical noncommutative spaces  are string like and this is a good motivation to study physics in these spaces. In this paper we study some interesting and important quantum systems in DNCS.\\
\section{The harmonic oscillator}
Harmonic oscillator is an important model in many branches of physics. It can be used to illustrate the basic concepts and methods in quantum mechanics. It also has applications in a variety of branches of modern physics including spectroscopy, condensed matter physics, nuclear structure, quantum optics, quantum field theory and statistical mechanics.\\
The Hamiltonian of a two dimensional harmonic oscillator in DNCS is given by:
\begin{equation}
H(x,y,p_x,p_y)=\dfrac{p_x^2+p_y^2 }{2m}+\dfrac{1}{2}m\omega^2(x^2+y^2 )
\end{equation}
Using Eq.(\ref{p6}), we rewrite the Hamiltonian in terms of $\theta$-noncommutative variables:
 \begin{align}\label{fg1}
h(x_0,y_0,p_{x_0},p_{y_0})&=\dfrac{1}{2m}[p_{x_0}^2+(1+\tau y_0^2)^{1/2}p_{y_0}(1+\tau y_0^2)p_{y_0}(1+\tau y_0^2)^{1/2}]\nonumber \\& + \dfrac{1}{2}m\omega^2[(1+\tau y_0^2)^{1/2}x_0(1+\tau y_0^2)x_0(1+\tau y_0^2)^{1/2}+y_0^2]
\end{align}
with the help of the Bopp-shift this Hamiltonian can be stated in terms of the standard  commutative variables. To the first order in  $\theta$ and $\tau$ we have: 
\begin{align}
H_{ho}(x_s,y_s,p_{x_s},p_{y_s})&=\dfrac{p_{x_s }^2+p_{y_s }^2 }{2m}+\dfrac{1}{2}m\omega^2(x_s^2+y_s^2 )- \dfrac{m\omega^2\theta}{2\hbar}L_{z_s}\nonumber\\
&+\dfrac{\tau}{m}y_s^2p_{y_s}^2+m\omega^2\tau y_s^2x_s^2-2i\dfrac{\tau\hbar}{m}y_sp_{y_s}-\dfrac{\tau\hbar^2}{2m}\nonumber\\&=H^s_{ho}+ H_{\theta ,\tau}
\end{align}
where $H^s_{ho}$ is the Hamiltonian of harmonic oscillator in commutative spaces. If we set $\tau=0$  we get the Hamiltonian of  harmonic oscillator in   $\theta$-noncommutative case i.e. Eq.(8) of  Ref.[5](we note that  in this equation  to the first order in $\theta$, the paramete $k$=1).\\
 Since the noncommutativity parameters $\tau$ and $\theta$ if they are  non-zero  should be very small, we use perturbation theory to find the spectrum of  quantum systems.  The perturbation Hamiltonian can be rewritten as follows:
\begin{align}
H_{\theta ,\tau}&=- \dfrac{m\omega^2\theta}{2\hbar}L_{z_s}+\dfrac{\tau}{m}y_s^2p_{y_s}^2+m\omega^2\tau y_s^2x_s^2-2i\dfrac{\tau\hbar}{m}y_sp_{y_s}-\dfrac{\tau\hbar^2}{2m}\nonumber\\&=v_1+v_2+v_3+v_4+v_5
\end{align}
The contributions of the different parts of the perturbation  Hamiltonian are as follows:
\begin{align*}
\langle{n_x,n_y}|v_1|n_x',n_y'\rangle&=-\frac{i}{2}m\omega^2\theta\sqrt{(n_y'+1)n_x'}\delta_{n_x,n_x'+1}\delta_{n_y,n_y'-1}\nonumber \\&+\frac{i}{2}m\omega^2\theta\sqrt{(n_x'+1)n_y'}\delta_{n_x,n_x'-1}\delta_{n_y,n_y'+1} 
\nonumber\\
\langle{n_x,n_y}|v_2|n_x',n_y'\rangle&=\frac{\tau
\hbar^2}{4m}(2n_y'^2+2n_y'-1)\delta_{n_x,n_x'}\delta_{n_y,n_y'}\nonumber\\
\langle{n_x,n_y}|v_3|n_x',n_y'\rangle&=\frac{\tau\hbar^2}{4m}\Big[\sqrt{(n_y'-1)n_y'}\delta_{n_y,n_y'-2}+(2n_y'+1)\delta_{n_y,n_y'}\nonumber\\&+\sqrt{(n_y'+1)(n_y'+2)}\delta_{n_y,n_y'+2}\Big] \Big[\sqrt{(n_x'-1)n_x'}\delta_{n_x,n_x'-2}\nonumber\\&+(2n_x'+1)\delta_{n_x,n_x'}+\sqrt{(n_x'+1)(n_x'+2)}\delta_{n_x,n_x'+2}\Big]
\nonumber\\
\langle{n_x,n_y}|v_4|n_x',n_y'\rangle&=\frac{\tau\hbar^2}{m}\delta_{n_x,n_x'}\delta_{n_y,n_y'}
\end{align*} 
\begin{align}
\langle{n_x,n_y}|v_5|n_x',n_y'\rangle&=-\frac{\tau\hbar^2}{2m}\delta_{n_x,n_x'}\delta_{n_y,n_y'}
\end{align} 
So the elementes of the perturbation Hamiltonian are given by:

\begin{equation}
H^{\theta,\tau}_{ij}=\left\{ \begin{array}{rl}
&\dfrac{\tau \hbar^2}{4m}\sqrt{(j - 1)(j - 2)(g - j + 2)(g - j +
 3)}\ \ \ \ \ \ ;  \ \ i-j=-2 \\
\\
& \dfrac{i m \theta \omega^2}{2}\sqrt{(j - 1)(g - i + 1)}\ \ \ \ \ \ \ \  \ \ \ \ \ \ \ \ \ \ \ \ \ \ \ \ \ \
 ;  \ i-j=-1 \\
 \\
  &\dfrac{\tau \hbar^2}{4m}[ -2(j - 1)^2 + 2(j - 1) + 2(2j - 1)g +
  2]\ \ \ ;  \ i=j\\
 \\
& -\dfrac{i m\theta \omega^2}{2}\sqrt{(i - 1)(g - j + 1)}\ \ \ \ \ \ \ \ \  \ \ \ \ \ \  \ \ \ \ \ \ \ \ \
 ; \  i-j=1\\
\\
& \dfrac{\tau \hbar^2}{4m} \sqrt{j(j + 1)(g - j + 1)(g - j)}\ \ \ \ \  \ \ \ \ \ \ \ \ \ \ \ \ \
 \ ;  \ i-j=2
\end{array}\right.
\end{equation}\vspace{.2mm}
where $g=n_x+n_y$  and  integers $ i$  and $ j $  represent the row and column indices, respectively.\\
The energy shift for the ground state $n_x+n_y=0$ is:
\begin{equation}
 \Delta{E}=-\dfrac{\tau\hbar^2}{2m}
\end{equation}
There is an interesting point involved here, although the correction to the ground state of the harmonic oscillator due to the noncommutativity of space  to the first order in noncommutativity parameter vanishes in  non-dynamical spaces ( $\theta$-noncommutative space )[5], it has nonvanishing value $-\dfrac{\tau\hbar^2}{2m}$ in dynamical case.  The first excited state $n_x+n_y=1$, has two fold degeneracy. The perturbation matrix is given by:
 \begin{equation}
  \begin{bmatrix}
\dfrac{\tau\hbar^2}{m}   &   \dfrac{1}{2}m\omega^2\theta\\
-\dfrac{1}{2}m\omega^2\theta   &     2\dfrac{\tau\hbar^2}{m}
\end{bmatrix} 
 \end{equation}
and the corresponding  energies are:
\begin{equation}
E_{n_x+n_y=1}=\left\{ \begin{array}{rl}
2 \hbar \omega +\dfrac{3\tau\hbar^2}{2m}-\sqrt{\dfrac{\tau ^2\hbar^4}{4m^2}+\dfrac{m^2\omega ^4\theta ^2}{4}} \\                                             2 \hbar \omega +\dfrac{3\tau\hbar^2}{2m}+\sqrt{\dfrac{\tau ^2\hbar^4}{4m^2}+\dfrac{m^2\omega ^4\theta ^2}{4}}
  \end{array}\right.\vspace{5mm}
\end{equation}
 
\section{Quantum Hall effect}\label{g001}
Quantum Hall effect( QHE ) is the remakable observation of quantized transport in two dimensional electron gasses placed in a transverse magnetic field. Let us consider a moving particle with charge $q$ and mass $\mu$ in a two dimensional $x-y$ plane and submitted to a uniform magnetic field $\vec{B}$ in the $z$-direction. The components of vector potential can be taken as follows:
\begin{equation}
A_x=-\dfrac{1}{2}By\ \ \ \ \ \ \ , \ \ \ \ \ \ \ A_y=\dfrac{1}{2}Bx\ \ \ \ \ \ \ , \ \ \ \ \ \ \  A_z=0 
\end{equation}
For the Hamiltonian of the system we have:
\begin{align}
H(x,y,p_x,p_y)&=\dfrac{1}{2\mu}[(p_x+\dfrac{qB}{2 c}y)^2+(p_y-\dfrac{qB}{2 c}x)^2 ]
\end{align}
In terms of the $\theta$-noncommutative variables, the Hamiltonian takes the following form:
\begin{align}
H(x_0,y_0,p_{x_0},p_{y_0})&=\dfrac{1}{2\mu}\big[(p_{x_0}+\dfrac{qB}{2 c}y_0)^2+[(1+\tau y_0^2)^{1/2}p_{y_0}(1+\tau y_0^2)^{1/2}\nonumber \\& -\dfrac{qB}{2 c}(1+\tau y_0^2)^{1/2}x_0 (1+\tau y_0^2)^{1/2}]^2\big ]\nonumber \\&=h_{ho}^{\theta}-2\theta \mu\omega_L\tau^2\hbar y_0^2+2i\omega_L(\tau y_0+\tau^2y_0^3)[\hbar x_0+\theta p_{y_0}]\nonumber \\&-\omega_L \tau y_0^2p_{y_0}x_0+\tau \hbar \omega_L\theta -\omega_L l_{z_0} 
\end{align}
where $h_{ho}^{\theta}$ is the Hamiltonian of the harmonic oscillator in $\theta$-noncommutative space, with angular frequency $\omega_L=\dfrac{qB}{2\mu c}$. In terms of the standard commutative space variables it is given by:
\begin{align}
H(x_s,y_s,p_{x_s},p_{y_s})&=\dfrac{1}{2\mu}(p_{x_s }^2+p_{y_s }^2)+\dfrac{1}{2}\mu \omega_L^2(x_s^2+y_s^2 )- \omega_L L_{z_s} \nonumber\\&-\dfrac{\mu\omega_L^2\theta}{2\hbar}L_{z_s}+\dfrac{\tau}{\mu}y_s^2p_{y_s}^2+\mu \omega_L^2\tau y_s^2x_s^2-2i\dfrac{\tau\hbar}{\mu}y_sp_{y_s}-\dfrac{\tau\hbar^2}{2\mu} \nonumber \\&+2i \omega_L \tau\hbar y_s x_s-\omega_L \tau y_s^2 p_{y_s} x_s+\dfrac{\omega_L\theta}{2\hbar}(p_{x_s }^2+p_{y_s }^2)\nonumber\\&=H^s_{ho}+H^{\theta ,\tau}
\end{align}
It is convenient to write the perturbation Hamiltonian in circular cylinderical coordinates:
\begin{align*}
H^{\theta,\tau}&=-\dfrac{\tau\hbar^2}{\mu}\{\rho^2\sin^4\varphi\dfrac{\partial^2}{\partial\rho^2}-m^2\sin^2\varphi\cos^2\varphi-2im\sin^3\varphi\cos\varphi \nonumber \\ &+2im\rho\sin^3\varphi \cos\varphi\dfrac{\partial}{\partial\rho}\}+\mu\omega_L^2\tau\rho^4\sin^2\varphi\cos^2\varphi+\{im\cos^2\varphi \nonumber \\ &+\rho\sin\varphi\cos\varphi\dfrac{\partial}{\partial\rho}\}-2\dfrac{\tau\hbar^2}{\mu}\{\rho\sin^2\varphi\dfrac{\partial}{\partial\rho}+im\sin\varphi\cos\varphi\}-\dfrac{\tau\hbar^2}{2\mu}
\end{align*}
\begin{align}
&+2i\omega_L\tau\hbar\rho^2\sin\varphi\cos\varphi+i\hbar\omega_L\tau\{\sin^3\varphi\cos\varphi\dfrac{\partial}{\partial\rho}+im\rho^3\sin^2\varphi\nonumber \\&\  \cos^2\varphi\}-\hbar\omega_L\theta\{\sin^2\varphi\dfrac{\partial^2}{\partial\rho^2}-\dfrac{m^2}{\rho^2}\cos^2\varphi-2i\dfrac{m}{\rho^2}\sin\varphi\cos\varphi\nonumber \\&+2i\dfrac{m}{\rho}\sin\varphi\cos\varphi\dfrac{\partial}{\partial\rho}+\dfrac{\cos^2\varphi}{\rho}\dfrac{\partial}{\partial\rho}\}
\end{align}
Using the wave function:
\begin{equation*}\label{fg00}
\Psi_{n_\rho, m}(\rho, \varphi)=\sqrt{\dfrac{(n_\rho)! \ \xi^{2|m|+2}}{\pi (n_\rho +|m|)!}} \rho ^{|m|}L^{|m|} _{n_\rho} {(\xi^2\rho^2)}  e^{im\varphi } e^{-1/2{\xi ^2 \rho ^2}}\ \ \ ; \ \ \ \ \xi^2=\dfrac{\mu\omega_L}{\hbar}
\end{equation*}
we calculate the energy spectrum of the ground state:
\begin{equation}
E_{m=0,n_\rho=0}=\hbar \omega_L+\dfrac{1}{4\mu \xi ^4 } [\tau \mu ^2 \omega_L ^2+\tau \xi ^4 \hbar ^2+2\hbar \mu \theta \omega_L \xi ^6 ] =0.25\dfrac{\tau \hbar^{2}}{\mu}+\dfrac{1}{2}\hbar\theta\omega_{L}\xi^{2}
\end{equation}  
For the first excited state we have:
\begin{equation}
E_{m=0,1;n_\rho=1}=3\hbar \omega_L+\left\{ \begin{array}{rl}
&\dfrac{1}{4\mu \xi ^4 } [7 \tau \mu ^2 \omega_L ^2+7 \tau \xi ^4 \hbar ^2+6 \hbar \mu \theta \omega_L \xi ^6 ]=1.75\dfrac{\tau \hbar^{2}}{\mu}+ 1.5\hbar\theta\omega_{L}\xi^{2}\\                                           
&  \dfrac{1}{8\mu \xi ^4 } [(6 \tau-\theta \xi ^4) \mu ^2 \omega_L ^2+6 \tau \xi ^4 \hbar ^2+ \hbar \mu  \omega_L \xi ^2(-\tau+4\theta \xi ^4)]=0.7\dfrac{\tau \hbar^{2}}{\mu}+\\
&0.5\hbar\theta\omega_{L}\xi^{2}-0.125\mu\theta\omega_{L}^{2}
  \end{array}\right.
\end{equation} \\
By setting  $\tau=0$, we obtaion the results of non-dynamical spaces, for instance Eqs.(22) and (23) give the same values for the energy of the  ground and first excited states  of  the  system as Eq.(20) in [6].
\section{Hydrogen atom}
In this section we study the Hydrogen atom in a two dimensional dynamical noncommutative spaces. Electronic bound states around charged impurities in two dimensional systems can be described in terms of a two dimensional  Hydrogen atom.\\
The Hamiltonian of the system is as follows:
\begin{equation}
H(x,y,p_x,p_y)=\dfrac{1}{2m}(p_x^2+p_y^2)-\dfrac{ze^2}{{ \sqrt{x^2+y^2} }}
\end{equation}
In terms of the $\theta$-noncommutative variables it takes the following form:
 \begin{align}
h(x_0,y_0,p_{x_0},p_{y_0})=&\dfrac{1}{2m}[p_{x_0}^2+(1+\tau y_0^2)^{1/2}p_{y_0}(1+\tau y_0^2)p_{y_0}(1+\tau y_0^2)^{1/2}]\nonumber \\& -{ze^2}[(1+\tau y_0^2)^{1/2}x_0(1+\tau y_0^2)x_0(1+\tau y_0^2)^{1/2}+y_0^2]^{-1/2}\nonumber \\=&\dfrac{1}{2m}[p_{x_0}^2+(1+\tau y_0^2)^2 p_{y_0}^2 ]-\dfrac{2 i\hbar\tau}{m} y_0(1+\tau y_0^2) p_{y_0}\nonumber\\&-\dfrac{ \tau^2 \hbar^2}{m} y_0^2-\dfrac{\tau\hbar^2}{2\mu} -{ze^2}[(1+\tau y_0^2)^2 x_0^2+4\ i\theta\tau y_0\nonumber\\& (1+\tau y_0^2)x_0-\tau \theta^2-2\tau^2 \theta^2 y_0^2 + y_0^2 ]^{-1/2}
\end{align}
In standard commutative space we have:
\begin{align}
H(x_s,y_s,p_{x_s},p_{y_s})&=\dfrac{1}{2m}(p_{x_s }^2+p_{y_s }^2)-\dfrac{ze^2}{ (x_s^2+y_s^2)^{1/2 }}+\dfrac{ze^2}{{ 2(x_s^2+y_s^2)^{3/2} }}\nonumber \\&\times[2\tau{x_s }^2{y_s }^2-\dfrac{\theta}{\hbar}L_{z_s}]+\dfrac{\tau}{m}y_s^2p_{y_s}^2-2i\dfrac{\tau\hbar}{m}y_sp_{y_s}-\dfrac{\tau\hbar^2}{2m}\nonumber \\&=H^s+H^{\theta,\tau}
\end{align}
If we take  $\tau=0$ we obtain $H_{\theta}=-\dfrac{ze^2 L_{z}\theta}{{ 2\hbar r^{3} }}$, which is in agreement with the  results of non-dynamical  case presented in Eq.(2.5)  of Ref.[7](note that in [7],$\theta_{ij}$ is defined through  $\theta_{ij}=\dfrac{1}{2}\epsilon_{ijk}\theta_k$  but we  have chosen $\theta_{ij}=\epsilon_{ijk}\theta_k$).\\
Using the wave functions of  two dimentional  Hydrogen atom [8], The energy spectrum for the ground state and the first degenerate  excited states are obtain:
\begin{equation}
\triangle{E} =0.125\dfrac{\tau \hbar ^2} {m}
\end{equation}
and
\begin{equation}
\triangle{E}=\left\{\begin{array}{rl}
& 0.875\dfrac{\tau\hbar^2}{2m}\ \ \ \ \ \ \ \ \ \ \ \ \ \ \ \ \ \ \ \ \ \ \ \ \\
&0.75\dfrac{\tau\hbar^2}{m}-\sqrt{0.06\dfrac{\tau ^2\hbar^4}{m ^2}+0.04 \dfrac{(ze^2)^2\theta ^2}{a_0^6}} \\
&0.75\dfrac{\tau\hbar^2}{m}+\sqrt{0.06\dfrac{\tau ^2\hbar^4}{m ^2}+0.04 \dfrac{(ze^2)^2\theta ^2}{a_0^6}}
\end{array}\right.
\end{equation}
\\
\section{The Zeeman effect}
The Zeeman effect is very important in applications such as nuclear magnetic resonance spectroscopy, electron spin resonance spectroscopy, magnetic resonance imaging ( MRI ) and M$\ddot{o}$ssbaure spectroscopy. To study the effects of dynamical noncommutativity on Zeeman effect we consider  the  relevant  Hamiltonian:
\begin{align}
H(x,y,p_x,p_y)&=\dfrac{1}{2m}(p_x^2+p_y^2)-\dfrac{ze^2}{{ \sqrt{x^2+y^2} }}-\omega L_z\nonumber \\ &
=\dfrac{1}{2m}(p_x^2+p_y^2)-\dfrac{ze^2}{{ \sqrt{x^2+y^2} }}-\omega[xp_y-yp_x]
\end{align}
In terms of the $\theta$-noncommutative variables it takes the following form:
\begin{align}
H(x_0,y_0,p_{x_0},p_{y_0})&=\dfrac{1}{2m}[p_{x_0}^2+(1+\tau y_0^2)^2 p_{y_0}^2 ]-\dfrac{2 i\hbar\tau}{m} y_0(1+\tau y_0^2) p_{y_0}\nonumber\\&-\dfrac{ \tau^2 \hbar^2}{m} y_0^2-\dfrac{\tau\hbar^2}{2m} -{ze^2}[(1+\tau y_0^2)^2 x_0^2+4\ i\theta\tau \nonumber\\&y_0(1+\tau y_0^2)x_0-\tau \theta^2-2\tau^2 \theta^2 y_0^2 + y_0^2 ]^{1/2}-\omega\nonumber\\&[(1+\tau y^2_0)^{1/2}x_0(1+\tau y^2_0)p_{y_0}(1+\tau y^2_0)^{1/2}-y_0p_{x_0}]
\end{align}
\vspace{5mm}
Using Bopp-shift one can write down the Hamiltonian in standard commutative space:
\begin{align}
H(x_s,y_s,p_{x_s},p_{y_s})&=\dfrac{1}{2m}(p_{x_s }^2+p_{y_s }^2)-\dfrac{ze^2}{ (x_s^2+y_s^2)^{1/2 }}+\dfrac{ze^2}{{ 2(x_s^2+y_s^2)^{3/2} }}\nonumber \\&\times[2\tau{x_s }^2{y_s }^2-\dfrac{\theta}{\hbar}L_{z_s}]+\dfrac{\tau}{m}y_s^2p_{y_s}^2-2i\dfrac{\tau\hbar}{m}y_sp_{y_s}-\dfrac{\tau\hbar^2}{2m}\nonumber \\& -\omega [x_sp_{y_s}-y_sp_{x_s}]+\dfrac{\theta}{2\hbar}\omega(p_{x_s }^2+p_{y_s }^2)\nonumber \\&-2\tau \omega y_s^2 p_{y_s} x_s+ i \hbar \omega \tau y_s x_s
\end{align}
 If we choose $\tau=0$,  Eq.(31) reduces to the Eqs.(24), (25) and (26) of  the  Ref.[9] for the case of  non-dynamical spaces(to the first order in magnetic field $B$).
The corrections to the energy for the ground state and  degenerate first excited state are given by:
\begin{equation}
\triangle{E}=2 \dfrac{\hbar \omega  \theta }{a_0^2} + 0.125 \dfrac{\tau \hbar^2}{m}
\end{equation}
and
\begin{equation}
\triangle{E}=\left\{ \begin{array}{rl}
& 0.22 \dfrac{ \hbar \omega  \theta}{ a_0^2} + 0.875 \dfrac{\tau \hbar^2}{m}\\
&\dfrac{0.5}{a_0^6 m^2}  [-a_0^4 \hbar m(-1.5 a_0^2\tau \hbar-0.4 m \theta \omega)-0.5 a_0^3 ( a_0^6 \hbar^4 m^2 \tau^2 \\ & +0.64 m^4 ( ze^2)^2 \theta^2-17.92  z e^2 a_0^5\hbar m^4 \tau \theta \omega)^{1/2} ]=\\
& 0.75 \dfrac{\tau \hbar^{2}}{m}+0.2 \dfrac{\hbar\theta\omega}{a_{0}^{2}}-0.5 [(\dfrac{\tau \hbar^{2}}{m})^{2}-17.92 \dfrac{\tau\hbar^{2}}{m}\dfrac{(z e^{4})\theta\omega m^{2}}{\hbar^{3}}+0.64 \dfrac{(z e^{2})^{2}\theta^{2}}{a_{0}^{6}} ]^{\dfrac{1}{2}}\\
&\dfrac{0.5}{a_0^6  m^2}  [-a_0^4 \hbar m(-1.5 a_0^2\tau \hbar-0.4 m \theta \omega)+0.5 a_0^3 ( a_0^6 \hbar^4  m^2 \tau^2 \\ & + 0.64  m^4 ( ze^2)^2 \theta^2-17.92  z e^2 a_0^5\hbar m^4 \tau \theta \omega )^{1/2}=\\
&  0.75 \dfrac{\tau \hbar^{2}}{m}+0.2 \dfrac{\hbar\theta\omega}{a_{0}^{2}}+0.5 [(\dfrac{\tau \hbar^{2}}{m})^{2}-17.92 \dfrac{\tau\hbar^{2}}{m}\dfrac{(z e^{4})\theta\omega m^{2}}{\hbar^{3}}+0.64 \dfrac{(z e^{2})^{2}\theta^{2}}{a_{0}^{6}} ]^{\dfrac{1}{2}} 
\end{array}\right.
\end{equation}
where  $\omega=\dfrac{eB}{2\mu c}$ and  $a_0$ is the Bohr radius.\\
\section{The Stark effect}
Stark effect has become an increasingly important part of atomic, molecular and optical physics.
To study the effects of the external electric field on the spectrum of the Hydrogen atom in dynamical  noncommutative spaces, we consider the electric field in the $x$-direction. The Hamiltonian is
\begin{equation}
H(x,y,p_x,p_y)=\dfrac{1}{2m}(p_x^2+p_y^2)-\dfrac{ze^2}{{ \sqrt{x^2+y^2} }}+eEx
\end{equation}
For the ground state we have :
\begin{equation}
\triangle{E} =0.125\dfrac{\tau \hbar ^2} {m}
\end{equation}
It can be  shown that  for the first excited state the energy correction to the first order is also proportional to  $\dfrac{\tau \hbar ^2} {m}$.\\
The same results are obtained if the electric field applied in the $y$-direction.\\
It is shown in [7] that at tree level the contribution to the Stark effect is zero in non-dynamical case  but we observe that in dynamical noncommutative spaces the contribution to the Stark effect is nonzero, therefore the Stark effect can be employed  to determine whether the noncommutativity of space is dynamical or nondynamical.  it is instructive to understand the mechanism intuitively. As we already mentioned the energy correction due to the external electric field on the energy spectrum of the Hydrogen  ground state in dynamical noncommutative space is not zero. This means that there is a permanent electric  dipole for the ground state of the  hydrogen atom in DNCS. To understand how this electric dipole is produced, we note that the fundamental objects in DNCS are not point like but they are string like. On the other hand it is shown  in [10] that strings can act as electric dipoles, so the fundamental particles(strings) in dynamical noncommutative spaces  for instance  electrons, quarks (neurons and proto
ns)   can have electric dipoles  due to dynamical noncommutativity of space which leads to permanent  electic dipole  for the ground state of the Hydrogen atom.\\

\section{Conclusions}

We have studied some fundamental and interesting quantum systems in dynamical noncommutative spaces. The energy shift for the ground state of the harmonic oscillator due to noncommutativity of space is zero for non-dynamical case while it has non-vanishing value in DNCS.  We have shown  that Stark effect can be used to check whether the noncommutativity of space is dynamical or non-dynamical. It seems that  in the dynamical noncommutative space  introduced here, there is a fundamental energy so that the corrections due to dynamical noncommutativity on the energy of a quantum system can be stated  in terms of  $g\dfrac{\tau\hbar^{2}}{m}$ in which $\dfrac{\tau\hbar^{2}}{m}$ is fundamental enegy independent of the system and the factor $g$ depends on the  kind of  the system. For instance, for the ground state of the harmonic oscillator and quantum Hall effect, the values of g are $-0.5$ and $0.25$ respectively. For  Hydrogen atom, Zeeman and stark effects it is $0.125$.    
We confirm that in the limit $\tau \rightarrow 0 $, we obtain the results presented in previous works devoted to non-dynamical noncommutative spaces. Finally one can use the energy accuracy  measurement $10^{-12} eV$ [11] to  impose some bounds on the value of the noncommutativity parameter $\tau$ : $\dfrac{\tau\hbar^{2}}{m}<10^{-12} eV$ .

\end{document}